\documentclass[a4paper,11pt]{article}
\pdfoutput=1
\usepackage{jheppub}
\usepackage{graphicx}
\usepackage{amsmath}
\usepackage{amsfonts}
\usepackage{mathrsfs}
\usepackage{amssymb}
\usepackage{slashed}
\usepackage[colorlinks=true,linkcolor=blue]{hyperref}
\usepackage[x11names]{xcolor}
\usepackage{bm}
\usepackage{empheq}
\usepackage[super]{nth}

\def\Eq#1{Eq.~\eqref{#1}}

\def\Dilog{\mathrm{Li}_2}
\def\CC{\mathcal{C}}
\def\OO{\mathcal{O}}
\def\mD{m_{\mathrm{D}}}
\def\nf{n_{\mathrm{f}}}
\def\ffhc{\mathrm{f\bar{f}hc}}
\def\ffh{\mathrm{f\bar{f}h}}
\def\fh{\mathrm{fh}}

\def\Qmax{\mathcal{Q}_{\mathrm{max}}}

\title{Hot and Dense QCD Shear Viscosity at Leading Log}
\author{Isabella Danhoni, Guy D. Moore}
\affiliation{Institut f\"ur Kernphysik, Technische Universit\"at Darmstadt\\
Schlossgartenstra{\ss}e 2, D-64289 Darmstadt, Germany}
\emailAdd{idanhoni@theorie.ikp.physik.tu-darmstadt.de,guy.moore@physik.tu-darmstadt.de}

\abstract{
 The leading-order weak-coupling shear viscosity of QCD was computed almost 20 years ago, and the extension to next-to-leading order is 4 years old.
 But these results have never been applied at finite baryon chemical potential $\mu$, despite the fact that intermediate-energy heavy ion collisions and merging neutron stars may explore the Quark-Gluon Plasma in a regime where baryon chemical potentials are large.
 Here we extend the leading-log shear viscosity calculation to finite $\mu$, and we argue that the convergence of the weak-coupling expansion, while questionable for achievable plasmas, should be better at $\mu > T$ than at $\mu=0$.

}


\date{\today}

\begin{document}

\maketitle

\section{Introduction}
\label{sec:intro}
The quark-gluon plasma (QGP) is the deconfined phase of the strongly interacting QCD. This phase existed in the early universe and is also briefly produced in heavy-ion collisions, and has been intensively studied in the past few decades\cite{Grefa:2022sav}.
Experimental results, starting with the early days of the Relativistic Heavy-Ion Collider (RHIC), indicate that QGP is a strongly coupled fluid\cite{PHENIX:2004vcz,STAR:2005gfr,ALICE:2010suc} and behaves very closely to a perfect fluid \cite{Heinz:2013th,Luzum:2013yya,JETSCAPE:2020mzn}.
This is in contrast to the expected behavior at extremely high temperatures (which existed early in the Universe but cannot be achieved in heavy ion collisions).
In this regime, the running coupling becomes small enough that a perturbative treatment may be valid.
For thermodynamical quantities, it appears that suitably resummed perturbative calculations work starting at a few times the deconfinement temperature \cite{Laine:2003ay,Hietanen:2008tv}.

\begin{figure}[hbt]
    \centering
    \includegraphics[scale=0.90]{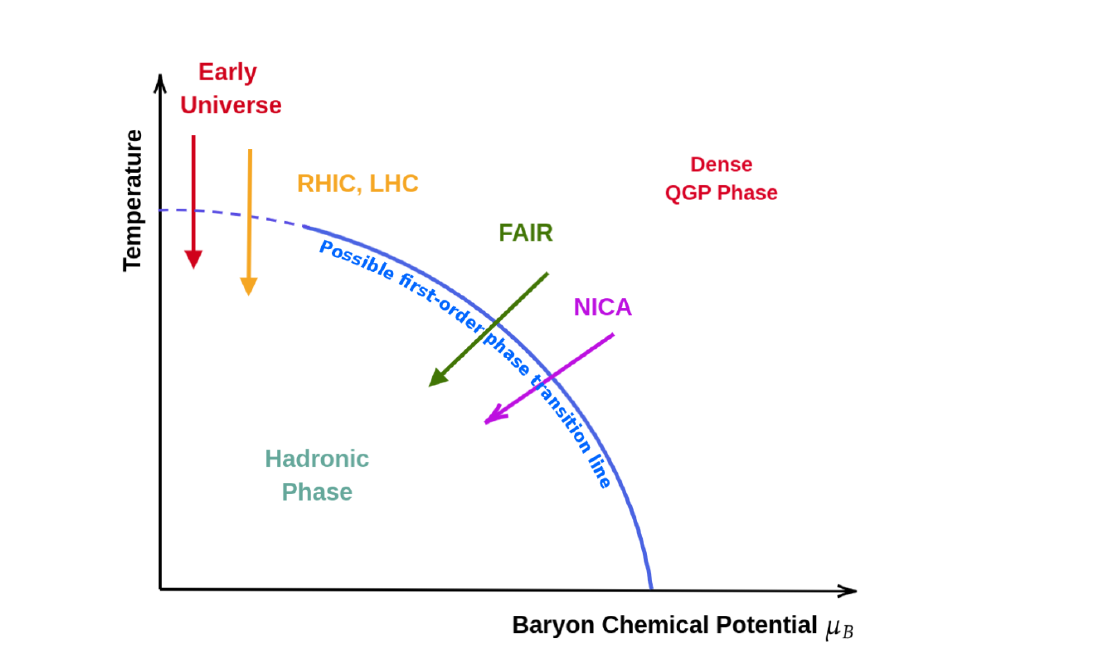}
    \caption{QCD phase diagram. The yellow line indicates the region that can be accessed by RHIC and LHC experiments. The green and purple lines show the region accessed by FAIR and NICA, respectively. In red, we indicate the two regions discussed in this introduction. The early universe in the $\mu_B=0$ axis, and the dense QGP region, where lies the interest of this paper. }
    \label{qcd}
\end{figure}

At ultra-relativistic energies at the Large Hadron Collider (LHC) or RHIC, QGP is produced with nearly zero baryon chemical potential \cite{Soloveva:2021pdd}.
Under these conditions, lattice gauge theory calculations are applicable, and the thermodynamic behavior is very well studied.
In comparison, lower-energy collisions generate a quark-gluon plasma with significant net baryon number, such that the quark chemical potential is several times higher than the temperature.
This regime cannot be directly studied on the lattice and is therefore much less well understood.
There are experimental efforts such as RHIC-BES \cite{STAR:2014egu}, HADES\cite{HADES:2019auv},
FAIR\cite{Friese:2006dj} and NICA\cite{Kekelidze:2017tgp} that are or will be focused in bringing some clarity over this region of the QCD phase diagram.
But as always, a vigorous theoretical program is needed in parallel.

In exploring the high-temperature regime without chemical potentials, it has often been useful to understand the behavior at still-higher temperatures, where perturbative methods become available.
For instance, although the perturbative calculation of the shear viscosity is, properly speaking, not applicable to the temperatures achieved in experiment, its knowledge has provided a useful ``boundary condition'' for the behavior in the physically relevant region, see for instance Ref.~\cite{Csernai:2006zz}.
Therefore we deem it valuable to explore the regime of high temperature with large 
quark chemical potentials, for instance, $T$ larger than the QCD scale and $\mu$ at least a few times $T$.
One can then use the determined behavior as a limiting or boundary behavior for the physically relevant regime, exploring, for instance, whether one expects a similarly small shear viscosity to the low-$\mu$ case, or a larger value for shear viscosity.

Extrapolating from high temperatures to the physically achieved temperatures at high density is admittedly a stretch.
But we argue that, at least at high temperatures, the perturbative series should work better at high chemical potential $\mu > T > T_c$ than is the case along the temperature axis.
The physics which ruins the perturbative expansion at $\mu=0$ is the importance and strong-coupling behavior of ``soft'' ($p \lesssim T$) gluons.
The contribution of such gluons is enhanced by Bose stimulation factors
\cite{Linde:1980ts} and they lead to $\OO(g^2 N_c T / \mD)$ next-to-leading order (NLO) contributions in the collision operator
\cite{Caron-Huot:2008zna} which can be larger than the leading-order contribution and imperil the series convergence even at temperatures many times higher than the QCD scale
\cite{Ghiglieri:2018dib}.
However, at high chemical potential, scattering from quarks is enhanced by a factor of order $\mu^2 / T^2$.
This is reflected in a larger value for the Debye screening mass $\mD \sim g \mu$ rather than $\mD \sim gT$ in the absence of chemical potentials.
As gluons play less of a role, the importance of their $\OO(g^2 N_c T/\mD$) NLO contribution is smaller, and the perturbative expansion is better behaved.

With this motivation, we will take the first step towards a comprehensive study of the influence of chemical potentials on the QCD shear viscosity.
Namely, we extend the Arnold-Moore-Yaffe (AMY) calculations for shear viscosity at leading log
\cite{Arnold:2000dr}
in weakly coupled high-temperature and dense QCD.
``High temperature'' in this work means the same as in \cite{Arnold:2000dr,Arnold:2003zc}, that the temperature is much larger than the zero-temperature masses of elementary particles, \textit{i.e.}, we require that $T \gg \Lambda_{QCD}$ and $T \gg m_q$, and dense means that $\mu \geq T$.
In section 2, we review the kinetic theory that was used by Arnold, Moore and Yaffe in \cite{Arnold:2000dr}.
First, we introduce the necessary concepts, then we show how to obtain shear viscosity using a variational method to solve the linearized Boltzmann equation.

For the leading log order, one can perform the collision integrals using some simplifications, as shown in \cite{Arnold:2000dr}. In section 3, we quickly comment on these simplifications and show how one can perform the integrals of the relevant Feynman diagrams at high $\mu$. The analysis of our results can be seen in sections 4 and 5. Our calculations show that for hot dense QCD, meaning that $\mu^2 \gg T^2$, the shear viscosity to entropy ratio scales as $\eta/s \propto \mu^2 / T^2$.

\section{Review of Viscosity and Kinetic Theory}
\label{sec:kin}

To efficiently study transport coefficients in a weakly-coupled, hot dense plasma, one must start by constructing a kinetic theory that reproduces the dynamics of the quantum field theory at the level of precision required.
After that, one can apply this kinetic theory to the processes of interest \cite{Arnold:2002zm}. To compute shear viscosity, it will be sufficient to solve Boltzmann equations linearized in small deviations away from an equilibrium state of given temperature T and chemical potential $\mu$.

\subsection{Setup}
\label{sec:setup}

In a kinetic description the properties of a system, such as expectation values of components of the stress tensor, are dominated by the 2-point function, which is characterized by a statistical distribution function $f(\vec p,\vec x, t)$.
(For a review of kinetic theory in QCD, see for instance \cite{Kadanoff,Arnold:1998cy}.)
Shear viscosity is a property relevant for systems with space-nonuniform flow velocity, meaning that the local equilibrium form of the statistical function is:%
\footnote{We use a mostly-positive metric tensor $g_{\mu\nu} = \mathrm{Diag}[-1,+1,+1,+1]$,
and natural units, $c=1$, $\hbar = 1$. Capital letters $P,K$ are 4-vectors, lower case letters $\vec p,\vec k$ or $p_i,k_i$ are the space components, $p^0,k^0$ are the time components, and $p,k$ are the magnitudes of the space components.}
\begin{equation}
\label{f0}
    f_0^a(\vec p,\vec x,t) = \frac{1}{\exp(\gamma \beta ( p^0 - u_i p^i - \mu_a ) \mp 1 }
    = \frac{1}{\exp (\beta (-u_\mu P^\mu - \mu_a)) \mp 1}.
\end{equation}
Here $u^\mu = (\gamma , \gamma \vec v)$ is the local velocity 4-vector, $\gamma$ is the associated relativistic gamma-factor, and $\beta=\beta(\vec x,t) = 1/T$ is the local temperature.
The sign $\mp 1$ is $-$ for bosons and $+$ for fermions (the upper sign will always apply to bosons), and $\mu_a$ is the chemical potential for species $a$, which will be nonzero for quark species and opposite for a particle and its antiparticle.
In this work we will consider $SU(3)$ gauge theory with $\nf$ vectorlike quark species each with the same chemical potential.

If $\beta$ varies in space, $\partial_i \beta \neq 0$, or if the flow has divergence, $\partial_i u_i\neq 0$, then $f_0^a$ will be time dependent.
We shall ignore these cases in this paper and concentrate instead on the case where the flow velocity has a traceless-symmetric spatial derivative.
Choosing a frame such that $u_i(\vec x=0) = 0$, we consider the case in which
\begin{equation}
\label{sigma}
    \sigma_{ij} \equiv \partial_i u_j + \partial_j u_i - \frac{2}{3} \delta_{ij} \partial_k u_k \neq 0.
\end{equation}
$\sigma_{ij}$ is called the shear tensor.
In this case, one expects that the system will leave equilibrium, and we will write the nonequilibrium form of the statistical function as:
\begin{equation}
\label{f1}
    f^a(\vec p,\vec x) = f_0^a(\vec p, \vec x) + f_0(1 \pm f_0)\, f_1^a(\vec p,\vec x) \,,
\end{equation}
where we choose to split $f^a$ into an equilibrium and nonequilibrium correlator according to the Landau-Lifshitz conventions under which $f_0$ contributes all of the energy and momentum density in the system.
The choice to normalize $f_1$ with a factor of $f_0(1\pm f_0)$ will be convenient in what follows.

Even though the term $f_0(1\pm f_0) f_1$ does not contribute to the energy density $T^{00}$ or momentum density $T^{0i}$, one expects it to contribute to the stress tensor.
For small and slowly varying $\sigma_{ij}$, that dependence should be linear, and one defines:
\begin{equation}
\label{etadef}
    T_{ij} = P g_{ij} - \eta \, \sigma_{ij}
\end{equation}
with $P$ the pressure
and $\eta$, the shear viscosity, defined as the coefficient of the linear response of $T_{ij}$ to nonvanishing $\sigma_{ij}$.
If the fluid flow also has a divergence $\theta = \partial_i u_i$, then an additional term $T_{ij} = \ldots - \zeta \theta g_{ij}$ would be present, with $\zeta$ the bulk viscosity.
We will not attempt to study $\zeta$ here, focusing instead on $\eta$.

The stress tensor is determined from $f^a(\vec p,\vec x)$, at leading order and neglecting particle masses, through
\begin{equation}
\label{Tij}
    T_{ij}(\vec x) = \sum_a^{\ffhc} \int \frac{d^3 p}{(2\pi)^3} \frac{p_i p_j}{p} f^a(\vec p,\vec x)
\end{equation}
where the sum runs over all colors $\mathrm c$, helicities $\mathrm h$, flavors/particle types $\mathrm{f}$, and particle/antiparticle $\bar{\mathrm f}$.
So the sum has 16 terms due to gluons and $12\nf$ terms due to $\nf$ flavors of quarks.

``All'' that remains is to determine $f_1^a$ from $\sigma_{ij}$.
In a weakly coupled system the dynamics of $f^a$ are determined from a Boltzmann equation:
\begin{equation}
    \Big[\frac{\partial}{\partial t} + \vec{v}_p \cdot \frac{\partial}{\partial \vec{x}} + \vec{F}_{\mathrm{ext}} \cdot \frac{\partial}{\partial \vec{p}}\Big] f^a(\vec{p},\vec{x},t) = - \CC^a[f].
    \label{boltz}
\end{equation}
Neither the time derivative nor the external force will be relevant for viscosity calculations. 
Neglecting thermal and Lagrangian masses, the velocity vector is the unit vector in the direction of the momentum, $\vec{v}_p = \hat{p} = \vec p / p$.
For leading-log calculations it is sufficient to consider $2\leftrightarrow 2$ collisions.
The collision operator can be written as:
\begin{align}
    \nonumber
    \CC^a[f](\vec{p}) =\frac{1}{2} \sum_{bcd} \int_{\vec{k},\vec{p},\vec{k'}} &
    \frac{|\mathcal{M}_{abcd}(P,K,P',K')|^2}{2p^0\, 2k^0\, 2p'{}^0\, 2k'{}^0}
    (2\pi)^4 \delta^4(P+K-P'-K') \\
    \nonumber
    &\times \Big\{ f^a(\vec{p})f^b(\vec{k})[1\pm f^c(\vec{p'})][1\pm f^d(\vec{k'})] \\
    & \phantom{\times \Big\{ } {} -  f^c(\vec{p'})f^d(\vec{k'})[1\pm f^a(\vec{p})][1\pm f^b(\vec{k})]\Big\}
    \label{collop22}
\end{align}
where the incoming/outgoing momenta $\vec p,\vec k$ and $\vec p',\vec k'$ are all on shell, $p^0=p$.
We write $\int_{\vec{k}} = \int \frac{d^3\vec{k}}{(2\pi)^3}$ to simplify the notation.
The factor $1/2$ avoids a double counting in the external state sum when $c\neq d$ and is the relevant symmetry factor when $c=d$.
Given an explicit expression for the matrix elements, it should be possible in principle to solve this Boltzmann equation.
The remainder of the paper will be dedicated to doing this at leading-log order within full QCD with chemical potentials.
But there are still a few definitions which are useful in setting up the problem.

Introducing a more convenient rescaling of $\sigma_{ij}$,
\begin{equation}
\label{Xdef}
    X_{ij} \equiv \frac{1}{\sqrt{6}} \sigma_{ij}
\end{equation}
one finds after a little work that the LHS of \Eq{boltz} can be written as:
\begin{align}
\label{LHSboltzmann}
    \hat p \cdot \partial_x f_0^a(\vec p,\vec x) & = \beta p^0 f_0^a (1 \pm f_0^a) X_{ij}(x) \, I_{ij}(\vec p)  \; \equiv \; \beta^2 X_{ij}(x) S^a_{ij}(\vec p) \,,
    \\
    \label{IijDef}
    I_{ij}(\vec p) & \equiv \sqrt{\frac 32} \left( \hat p_i \hat p_j - \frac{1}{3} \delta_{ij} \right) \,,
    \\
    \label{SijDef}
    S^a_{ij}(\vec p) & \equiv p^0 \, T \, f_0^a(1\pm f_0^a) I_{ij}(\vec p) \,.
\end{align}
The normalization is chosen such that $I_{ij} I_{ij} = 1$.
Because this expression is already linear in $X_{ij}$ when we use the equilibrium $f_0$, inserting $f_1$ will lead to an expression which is quadratic in $X_{ij}$ or involves derivatives of this quantity.
Such terms are not needed here (see Ref.~\cite{York:2008rr} for applications where this is needed).

The collision operator vanishes if we insert $f \to f_0$, so it will be linear in $f_1(\vec p)$, that is, $f_1(\vec p) \propto X_{ij}(\vec p)$.
Because $f_1$ is a scalar quantity, it must also be proportional to the contraction $f_1(\vec p,\vec x) \propto X_{ij}(\vec x) I_{ij}(\vec p)$.
Therefore, without loss of generality we can write the departure from equilibrium as
\begin{equation}
\label{chidef}
    f_1^a(\vec p,\vec x) = \beta^2 X_{ij}(\vec x) \chi^a_{ij}(\vec p) = \beta^2 X_{ij}(\vec x) I_{ij}(\vec p) \chi^a(p) \,.
\end{equation}
Here $\chi^a(p)$ is a species-dependent, pure scalar function of $p$ which characterizes the departure from equilibrium in the presence of shear stress.
Since both sides of \Eq{boltz} are proportional to $X_{ij}$, we can rewrite that equation as:
\begin{equation}
    S^a_{ij}(\vec p) = \CC \chi^a_{ij}(\vec p) \,.
    \label{boltz2}
\end{equation}
Here $\CC \chi^a_{ij}(\vec p)$ means the collision operator of \Eq{collop22} with
$f^a_1 \to \chi^a_{ij}(\vec p)$ inserted as the departure from equilibrium -- that is, the collision operator is guaranteed to be proportional to $X_{ij}$, and $\CC \chi^a_{ij}(p)$ is the collision operator with this factor of $\beta^2 X_{ij}$ stripped off.

\subsection{Variational solution}
\label{sec:variational}

The variational approach consists in converting \Eq{boltz2} into an equivalent variational problem.
This strategy was used in \cite{Arnold:2000dr,Arnold:2003zc} in leading log and leading order calculations.
One starts by defining the inner product:
 \begin{equation}
 \label{innerprod}
     (f,g) = \beta^3 \sum^{\mathrm{\ffhc}}_{a} \int_{\vec{p}} f^a(\vec{p})g^a(\vec{p}) .
 \end{equation}
It can be shown that the collision operator is hermitian with respect to this inner product.
We define
 \begin{equation}
 \label{Qdef}
     \mathcal{Q}[\chi] = \left(\chi_{ij}, S_{ij}\right)- \frac{1}{2}\left(\chi_{ij},\CC\chi_{ij}\right) \,,
 \end{equation}
which can be viewed as a functional over the space of $\chi^a(p)$ values.
Setting the variation with respect to $\chi^a(p)$ equal to zero to find the extremum (maximum) of this functional returns \Eq{boltz2}.
That is, the functional $\mathcal{Q}$ takes its maximum value when the Boltzmann equation is satisfied.
Furthermore, its value at this maximum,
 \begin{equation}
 \label{Qmaxval}
     \Qmax=\frac{1}{2}(\chi_{ij}, S_{ij})=\frac{1}{2} (\chi_{ij},\CC\chi_{ij}) = \frac{1}{2} ( S_{ij},\CC^{-1} S_{ij} ) \,,
 \end{equation} 
determines the viscosity:
 \begin{equation}
 \label{etaQmax}
     \eta = \frac{2}{15}\Qmax.
 \end{equation}
The terms in this functional can be identified as the source:
 \begin{align}
    \nonumber
     (\chi_{ij}, S_{ij}) &= \beta^2 \sum^{\ffhc}_{a}\int_{\vec{p}}f_0(\vec{p})[1 \pm f_0(\vec{p})]|\vec{p}|\chi^a I_{ij}(\hat{p})I_{ij}(\hat{p})\\
     &= \beta^2 \sum^{\ffhc}_{a}\int_{\vec{p}}f_0(\vec{p})[1 \pm f_0(\vec{p})]|\vec{p}|\chi^a 
     \label{sourceint}
 \end{align}
and the collision integral, which after some symmetrization becomes: 
\begin{align}
 \nonumber
   (\chi_{ij},\CC\chi_{ij}) = \frac{\beta^3}{8} \sum^{\ffhc}_{abcd}\int_{\vec{p},\vec{k},\vec{p'},\vec{k'}} &
    \frac{|\mathcal{M}_{abcd}(P,K,P',K')|^2}{2p^0\, 2k^0\, 2p'{}^0\, 2k'{}^0}
   (2\pi)^4\delta^4(P+K-P'-K')\\
   \nonumber
   & \times  f^a_0(p)f^b_0(k)\left[1 \pm f^c_0(p')\right]\left[1 \pm f^d_0(k')\right] \\ &
   \times \left[\chi^a_{ij}(\vec{p}) + \chi^b_{ij}(\vec{k}) - \chi^c_{ij}(\vec{p'}) - \chi^d_{ij}(\vec{k'})\right]^2 .
   \label{Collexplicit}
 \end{align}
The sum is over all scattering processes in the plasma taking species $a$ and $b$ into species
$c$ and $d$.
The overall factor of 1/8 compensates for the eight times a given process is taken into account: $a\leftrightarrow b$, $c\leftrightarrow d$, and $(a,b)\leftrightarrow (c,d)$.

To maximize the functional $\Qmax$ exactly, one must work in the infinite-dimensional space of arbitrary functions $\chi(p)$.
However, as known from other variational problems, one can obtain highly accurate approximate results by performing a restricted extremization within a well-chosen finite dimensional subspace.
For this purpose, we expand the $\chi(p)$ functions into a finite basis set:
\begin{align}
\label{Ansatz}
  \chi^g(p)& =\sum_{m=1}^N a_m \phi^{(m)}(p), &
  \chi^q(p) & = \sum_{m=1}^N a_{m+N} \phi^{(m)}(p), &
  \chi^{\bar{q}}(p) & = \sum_{m=1}^N a_{m+2N} \phi^{(m)}(p) .  
\end{align}
Putting these back into the source and collision integrals, we get the source vector and the truncated scattering matrix:
\begin{align}
\label{SCbasis}
  (S_{ij},\chi_{ij}) & = \sum_{m} a_m \tilde{S}_m, &
  (\chi_{ij},\CC \chi_{ij}) & = \sum_{m,n} a_m \tilde{\CC}_{mn} a_n ,
\end{align}
where $\tilde S_m$, $\tilde \CC_{mn}$ are \Eq{sourceint} and \Eq{Collexplicit} with $\chi^a$ replaced by individual basis functions $\phi^{(m)}$, $\phi^{(n)}$, that is,
\begin{align}
\label{SCvalues}
    \tilde S_m & = ( \phi^{(m)} I_{ij} , S_{ij} ) \nonumber \\
    \tilde \CC_{mn} & = ( \phi^{(m)} I_{ij} , \CC \phi^{(n)} I_{ij} ) \,.
\end{align}
Within the subspace spanned by our basis, one estimates that:
\begin{align}
\label{Qbasis}
    \mathcal{Q}[\chi] & = a_m \tilde{S}_m - \frac{1}{2}a_m \tilde{\CC}_{mn} a_n  \\
      \eta & = \frac{1}{15}\tilde{S}_m \tilde{\CC}^{-1}_{mn} \tilde{S}_n \,,
    \label{visc3}
\end{align}
where $\tilde\CC^{-1}_{mn}$ is to be understood as the matrix inverse of $\tilde\CC_{mn}$, that is, $\tilde\CC^{-1}_{mn} \tilde\CC_{no} = \delta_{mo}$.

At this point, one has to choose a good basis set to the variational Ansatz.
We will use the set proposed in Ref.~\cite{Arnold:2000dr}:
\begin{equation}
\label{LGYbasis}
    \phi^{(m)}=\frac{p(p/T)^m}{(1+p/T)^{N-1}}, \hspace{0.3cm} m=1,...,N \,.
\end{equation}
The subspace spanned by this basis strictly increases as $N$ is increased.
The case $N=1$ (3 variational parameters, one each for $g,q,\bar{q}$) is known to work well
for the theory at $\mu=0$, but we will find that it works less well at, say, $\mu/T = 4$.
However, we will check that there is rapid convergence as we increase $N$.

Evaluating $\tilde S_m$ involves single integrals over known functions and is straightforward (numerically).
Evaluating the elements of $\tilde \CC$ is more involved.
We present it in the next section.

 \section{Collision Integrals at Finite Chemical Potentials }
\label{sec:coll}

In this section, our goal is to obtain the size of the contribution of each diagram to $(\chi_{ij},\CC\chi_{ij})$.
For that, we follow what was done by Arnold, Moore and Yaffe and shift the integration over $\vec{p}'$ into an integration over $\vec{q}=\vec{p}'-\vec{p}$.
At weak coupling, the diagrams are dominated by small $q$.
We work in the plasma frame, in spherical coordinates chosen with $\vec q$ as the $z$-axis and $\vec p$ lying in the $x-z$ plane:
\begin{align}
\label{collintunreduced}
  \nonumber
  (\chi_{ij},\CC\chi_{ij}) = & \frac{\beta^3}{(4\pi)^6}\sum^{\ffhc}_{abcd}\int_0^\infty q^2dq p^2dp k^2dk \int_{-1}^1 d(\cos{\theta_{pq}})d(\cos{\theta_{kq}})\int_0^{2\pi}d\phi\\
  \nonumber
  & \times \frac{|\mathcal{M}^{ab}_{cd}|^2}{pkp'k'}\delta^4(P+K-P'-K')f^a_0(p)f^b_0(k)[1 \pm f^a_0(p)][1 \pm f^b_0(k)]\\
  & \times \left[\chi^a_{ij}(\vec{p})+\chi^b_{ij}(\vec{k}) - \chi^c_{ij}(\vec{p'}) - \chi^d_{ij}(\vec{k'})\right]^2
\end{align}
The azimuthal angle of $\vec{k}$ is $\phi$ , and $\theta_{pq}$ is the plasma frame angle between $\vec{p}$ and $\vec{q}$. 

Following Baym et al. \cite{PhysRevLett.64.1867}, one introduces $\omega = p' - p$, which together with the energy-momentum delta functions allows one to further simplify the integrals:
\begin{align}
    \nonumber
  (\chi_{ij},\CC\chi_{ij}) = & \frac{\beta^3}{(4\pi)^6}\sum^{\ffhc}_{abcd}\int_0^\infty dq \int_{-q}^q d\omega \int_{\frac{q-\omega}{2}}^\infty dp\int_{\frac{q+\omega}{2}}^\infty dk\int_0^{2\pi}d\phi\\
  \nonumber
  & \times \frac{|\mathcal{M}^{ab}_{cd}|^2}{pkp'k'}ff^a_0(p)f^b_0(k)[1 \pm f^c_0(p')][1 \pm f^d_0(k')] \\
  & \times \left[\chi^a_{ij}(\vec{p})+\chi^b_{ij}(\vec{k}) - \chi^c_{ij}(\vec{p'}) - \chi^d_{ij}(\vec{k'})\right]^2 .
  \label{collint}
\end{align}
In general 
\begin{equation}
\label{Legendre}
\chi_{ij}^a(p) \chi_{ij}^b(k) = \chi^a(p) \chi^b(k) P_2(\cos \theta_{pk})
= \chi^a(p) \chi^b(k) \left( \frac 32 \cos^2 \theta_{pk} - \frac 12 \right)
\end{equation}
which depends on $\cos \theta_{pk}$ through the second Legendre polynomial $P_2(\cos\theta_{pk})$.
Therefore, to evaluate \Eq{collint} we need the cosines of the angles between $(p,k,p',k')$.

At weak coupling, scattering via a $t$-channel gluon has a screened IR divergence which leads to a large contribution at small $q$, see for instance \cite{PhysRevLett.64.1867} and references therein.
Similarly, $t$-channel quark exchange leads to log-large cross-sections which also change particle type $g \leftrightarrow q$ or $\bar{q}$.
The importance of each process is therefore enhanced by a logarithm of the ratio $\pi T / m$, with $\pi T$ the typical particle energy and $m$ a screening scale which is small in the weak-coupling limit.
A leading-log approximation consists of focusing only on these processes and the part of the scattering integrals where $m < q < \pi T$.
The resulting small-$q$ approximations will simplify our calculations, and hopefully still describe qualitatively how the viscosity changes as $\mu/T$ becomes larger.

The approximation $q \ll \pi T$ first allows us to continue the $p,k$ integrals in \Eq{collint} down to 0.
Next, it provides simplifications in evaluating matrix elements:
we can use the following approximations for the Mandelstam variables,
\begin{equation}
\label{smallQstu}
    \frac{-s}{t}\simeq \frac{u}{t} \simeq \frac{2pk}{q^2}(1-\cos{\phi}).
\end{equation}
Since $p \simeq p'$, the statistical functions only depend on $p,k$ -- that is, in evaluating $f_0$ we neglect the difference between $p$ and $p'$ and between $k'$ and $k$.
Finally, we can make some simplifications in treating the last line in \Eq{collint}.
For processes with a virtual quark in the $t$-channel we can approximate $\chi^a_{ij}(p) = \chi^a_{ij}(p')$.
For processes with a virtual gluon, the species labels $a,b$ are the same and this approximation gives zero.
We have to do better because the matrix element is more strongly divergent, so we expand to linear order in $q$:
\begin{equation}
\label{qexpand}
    \chi^a_{ij}(\vec{p}) - \chi^a_{ij}(\vec{p'}) = -\vec{q}\cdot \nabla \chi^a_{ij}(\vec{p}) + \mathcal{O}(q^2).
\end{equation}
\Eq{collint} requires the square of this quantity.
A simple evaluation (see Ref.~\cite{Arnold:2000dr}) gives:
 \begin{equation}
    [\chi^a_{ij}(\vec{p}) - \chi^a_{ij}(\vec{p'})]^2 = \omega^2 [\chi^a(|\vec{p}|)']^2+ 3 \frac{q^2-\omega^2}{p^2}[\chi^a(p)]^2 .
    \label{ang}
\end{equation}
The $\propto \omega^2$ term arises from the change in energy, while the $\propto q^2-\omega^2 = q_\perp^2$ term arises from the change in direction.
The coefficient is 3 because $I_{ij}$ represents an $\ell=2$ spherical harmonic; for a general $\ell$ the coefficient would be $\ell(\ell+1)/2$.
For the cross-contributions that arise in \Eq{collint} , one can use the approximations:
\begin{align}
\label{angleapprox1}
   & \cos{\theta_{pq}}\simeq\cos{\theta_{kq}}\simeq\cos{\theta_{p'q}}\simeq\cos{\theta_{k'q}}\simeq \frac{\omega}{q}\\
   \label{angleapprox2}
   & \cos{\theta_{pk}}\simeq\cos{\theta_{p'k}}\simeq\cos{\theta_{pk'}}\simeq\cos{\theta_{p'k'}}\simeq \frac{\omega^2}{q^2 }+ \frac{q^2-\omega^2}{q^2}\cos{\phi}.
\end{align}

\begin{figure}
    \centering
    \includegraphics[scale=0.3]{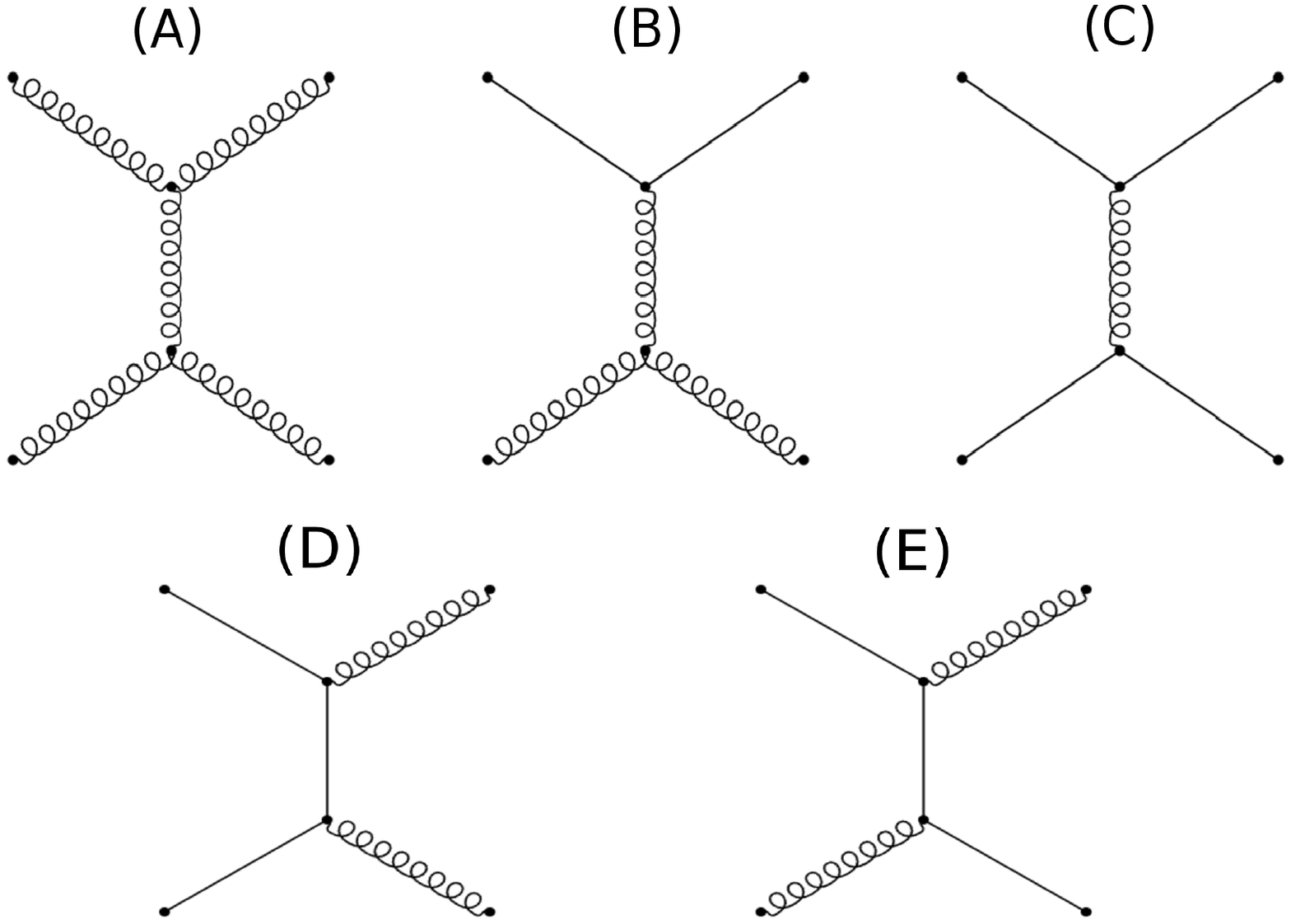}
    \caption{Leading log diagrams for the scattering processes in a gauge theory with fermions. Solid lines as fermions and the helices represent gluons. Time may be regarded as running horizontally. }
    \label{diag}
\end{figure}

At leading log order, there are five diagrams that must be computed, which are shown in figure \ref{diag}.
All approximations discussed so far can be used to compute the integrals for the five diagrams.
In the next subsections, we treat each diagram separately, compute the integrals and discuss the large $\mu$ region.

\subsection{Diagrams (A) and (B)}
\label{secAB}
Despite their differences, diagrams (A), (B), and (C) can be computed using some of the same approximations. As discussed in \cite{Arnold:2000dr}, in the soft exchange ($q \rightarrow 0$) limit, the vertices in diagrams (A)-(C) take on a universal form that depends only on the color charge of the particle that is scattering, which can be a gluon or a quark. Therefore, one can use this form of the vertex for the three diagrams: 
\begin{equation}
    2p^\mu g t^a\delta_{hh'}
\end{equation}
Where $t^a$ is the color generator, and $h$ and $h'$ represent the helicities of the scattering particles. This approximation makes the matrix element trivial to compute:
\begin{equation}
    |\mathcal{M}^{ab}_{cd}|^2_{\mathrm{lead-log}}=A_{ab}\frac{4p^2k^2}{q^4}(1-\cos{\phi})^2
\end{equation}
where $A_{ab}=4d_A T_{Ra}T_{Rb}g^4$, and we have only summed over the outgoing particles' helicity.
Here $d_A = N_c^2-1 = 8$ is the dimension of the adjoint representation and $T_R$ is the trace normalization, $1/2$ for quarks and $N_c=3$ for gluons.
Putting this matrix element back in the integral, and using the relation (\ref{ang}), the integrals over $\phi$ and $\omega$ are trivial. 
$\nabla_q \chi(p) \nabla_q \chi(k)$ cross-terms turn out to vanish after $\omega,\phi$ integration, leaving only terms where the statistical function arises twice on one side of the diagram.
The remaining ones are:
\begin{align}
    \nonumber
  (\chi_{ij},\CC\chi_{ij})^{(A)-(C)}=& \sum^{\ffh}_{ab}\frac{2A_{ab}\beta^3 8\pi}{(4\pi)^6}\int_0^\infty dq  \int_0^\infty dp\int_0^\infty dk \frac{k^2}{q}\\
  \nonumber
  & \times f^a_0(p)f^b_0(k)[1 \pm f^a_0(p)][1 \pm f^b_0(k)]\\
   & \times \left[p^2 [\chi^a(|\vec{p}|)']^2+ 6[\chi^a(p)]^2 \right]
\end{align}
The gluons shouldn't be affected by the chemical potential, so the integral over $k$ for diagram (A) can be computed with the relation:
\begin{equation}
    \int^\infty_0 dk k^2 f^b_0(k)[1 \pm f^b_0(k)]= \lambda_b T^3 \frac{\pi^2}{6}
\end{equation}
Where $\lambda_b$ is 2 for bosons and 1 for fermions.

On the other hand, for diagram (B), there are two contributions:
when both $\chi$ factors represent the fermion line, and when both $\chi$ factors represent the gluon line.
In the former case we can integrate over the gluon's energy, and in the latter case we can integrate over the quark energy.
From now on, we name the integrals after the departure from equilibrium we are left with, for example $B_f$ will be the result of integrating over the gluon momentum when quark $\chi$-factors are present.
Evaluating the contribution where both $\chi$ represent quarks, we find:
\begin{align}
    \nonumber
  (\chi_{ij},\CC\chi_{ij})^{(B_f)}=& \sum^{\fh}\frac{4d_A T_{Ra}T_{Rb}\lambda_b}{2^9 3\pi^3}  \int_0^\infty dp  f^f_0(p)\left[1 - f^f_0(p)\right]\\
  &\left\{p^2 \left[\chi^f(|\vec{p}|)'\right]^2+ 6\left[\chi^f(p)\right]^2\right\}.
\end{align}
Here $\sum^\mathrm{fh}$ means that the sum runs over flavor and helicity, but not over particle-antiparticle (which is explicit) or color (which is included in the color factors).
Similarly there is a contribution when the line represents an antiquark:
\begin{align}
    \nonumber
  (\chi_{ij},\CC\chi_{ij})^{(B_{\overline{f}})}=& \sum^{\fh}\frac{4d_A T_{Ra}T_{Rb}\lambda_b}{2^9 3\pi^3}  \int_0^\infty dp  f^{\overline{f}}_0(p)\left[1 - f^{\overline{f}}_0(p)\right]\\
  &\left\{p^2 \left[\chi^{\overline{f}}(|\vec{p}|)'\right]^2+ 6\left[\chi^{\overline{f}}(p)\right]^2\right\}.
\end{align}
These integrals depend on $\chi^f$ and must be performed numerically for each combination of basis functions.

To work on the departure from equilibrium of the bosons, we must start by integrating the fermion vertex.
The integral to be computed now is:
\begin{align}
    & \sum_\pm \int^\infty_0 dk \, k^2 \frac{1}{e^{\beta(k\pm \mu)}+1}\left[1 - \frac{1}{e^{\beta(k \pm \mu)}+1}\right]
    \nonumber \\
    = & \sum_\pm \int_0^\infty dk\, 2kT \frac{1}{e^{\beta (k\pm \mu)} + 1}
    \nonumber \\
    = & \frac{\pi^2 T^3}{3} + \mu^2 T \,.
\end{align}
Once we put the results back in \Eq{collint} we are left with:
\begin{align}
    (\chi_{ij},\CC\chi_{ij})^{(B_g)}=
    \sum^{\mathrm{h}}_{ab} & \frac{4A_{ab}\beta^3 8\pi}{(4\pi)^6}\left(\frac{T^3\pi^2}{6}+\frac{T\mu^2}{2}\right) \times 
    \\ \nonumber
    & \int_0^\infty dp \,
  f^g_0(p)[1+f^g_0(p)]\left\{p^2[\chi^g(p)']^2+6\chi^g(p)^2\right\} .
\end{align}
We see that the gluonic contribution to (B) is larger than the contribution to (A) in any regime where the $\mu^2$ term above is significantly larger than the $T^2$ term.
So at large chemical potentials, diagram (B) will be more important than diagram (A).

\subsection{Diagram (C)}
It is straightforward to compute diagram (C) using the same approaches that we develop above for (A) and (B).
Only fermionic departures from equilibrium are relevant, and the scatterer is always a fermion, leading to a $\mu^2 T + \pi^2 T^3/3$ type contribution like the one we saw in diagram (B).
Explicitly, one finds:
\begin{align}
    (\chi_{ij},\CC\chi_{ij})^{(C_f)}=
    \sum^{\fh}_{ab} & \frac{4A_{ab}\beta^3 8\pi}{(4\pi)^6}\left(\frac{T^3\pi^2}{6}+\frac{T\mu^2}{2}\right) \times \\ \nonumber
    & \int_0^\infty dp 
   \left\{f^f_0(p)[1 - f^f_0(p)]\left[p^2 [\chi^f(p)']^2+ 6[\chi^f(p)]^2\right]\right\}
 \\    \nonumber
    (\chi_{ij},\CC\chi_{ij})^{(C_{\overline{f}})}=
    \sum^{\overline{f}h}_{ab} & \frac{4A_{ab}\beta^3 8\pi}{(4\pi)^6}\left(\frac{T^3\pi^2}{6}+\frac{T\mu^2}{2}\right) \times \\
    \nonumber & \int_0^\infty dp
   \left\{f^{\overline{f}}_0(p)[1 - f^{\overline{f}}_0(p)]
  \left[p^2 [\chi^{\overline{f}}(p)']^2+ 6[\chi^{\overline{f}}(p)]^2\right]\right\}
\end{align}
Relative to the $\chi^f$-dependent term from diagram (B), this is enhanced by $\mu^2/T^2$ and will dominate in the high chemical-potential regime.
In this regime, the fermionic contributions to the shear viscosity dominate those from gluons because fermions dominate the stress tensor by a factor of $\mu^4/T^4$.
Therefore diagram (C) becomes the most important process in this limit.

NLO calculations
\cite{Ghiglieri:2018dib} show that the large corrections which imperil the convergence of the perturbative expansion arise from gluonic effects. 
Here we see that the scattering becomes quark-dominated in the large-$\mu$ regime, as shown in figure \ref{Q}, and these NLO effects should be suppressed in the large-$\mu$ case.  
Therefore we anticipate that perturbation theory should show better convergence in the large-$\mu$ regime than for the $\mu=0$ case.
Naturally, to really test this claim we will have to extend the current calculation to NLO.

\subsection{Diagrams (D) and (E)}
Diagrams (D) and (E) can be discussed together not only due to the similarities in their calculations, but also because both diagrams will be highly suppressed as the chemical potential increases.
The suppression for diagram (D) is a consequence of the fact that you need a fermion and an anti-fermion for this process to happen:
$f\overline{f}\rightarrow gg$, $\overline{f}f\rightarrow gg$, $gg\rightarrow f\overline{f}$ and $gg \rightarrow \overline{f}f$.
As the chemical potential increases, the number of anti-fermions decreases, and the rate for this process becomes small.

Using well-known results from QED, one should be able to easily compute diagram (D) and get:
\begin{equation}
|\mathcal{M}^{f\overline{f}}_{gg}|^2_{\mathrm{lead-log}}= A_f\Big(\frac{u}{t}+\frac{t}{u}\Big)
\end{equation}
where $A_f= 4d_A T_{Rf}C_{Rf}g^4$. Interchanging the outgoing legs makes $t/u \rightarrow u/t$, so one can keep only $u/t$ and multiply the result by 2. Since the divergence at leading log calculations is at most logarithmic, one may take all possible small q approximations $f_0(w+p)=f_0(p)$, $f_0(k-w)=f_0(k)$, $\cos{\theta_{pp'}}\simeq\cos{\theta_{kk'}}\simeq 1$. That includes the fact that at small q the matrix element can be rewritten as:
\begin{equation}
    \frac{u}{t}\simeq \frac{2kp}{q^2}(1-\cos{\phi}).
\end{equation}
Plugging these back into \Eq{collint} gives:
\begin{align}
  \nonumber
  (\chi_{ij},\CC\chi_{ij})^{(D)}=& \sum^{\overline{f}\fh}\frac{4A_f\beta^3}{(4\pi)^6}\int_0^\infty dq \int_{-q}^q dw \int_0^\infty dk \int_0^{2\pi}d\phi(1-\cos{\phi})\frac{2pk}{q^2}\\
  \nonumber
  &\times f^f_0(p)f^{\overline{f}}_0(k)[1 \pm f^g_0(p)][1 \pm f^g_0(k)]\Big\{[\chi^f(p)-\chi^g(k) ]^2\\
  \nonumber
  &+[\chi^{\overline{f}}(k)-\chi^g(k)]^2 +2P_2(\cos{\theta_{pk}})[\chi^f(p)-\chi^g(k)]\\
  &[\chi^{\overline{f}}(k)-\chi^g(k)]\Big\}.
\end{align}
The crossing terms vanish for $l > 0$ as a consequence of the orthogonality of the Legendre Polynomials. For the remaining terms, the integrals over $\omega$ and $\phi$ are trivial. One is left with two possible integrals, one over fermions and one over anti-fermions. For the term involving the fermions, one must compute:
\begin{eqnarray}
    \int dk \, k \frac{1}{e^{\beta(k-\mu)}+1}\left(1+\frac{1}{e^{\beta k}-1}\right)
\end{eqnarray}
And for the term involving anti-fermions:
\begin{align}
    \int dk \, k \frac{1}{e^{\beta(k+\mu)}+1}\left(1+\frac{1}{e^{\beta k}-1}\right)
    \label{D1}
\end{align}
The results of these two integrals give the contributions for diagram (D):
\begin{align}
  \nonumber
  (\chi_{ij},\CC\chi_{ij})^{(D_f)}=& \sum^{\fh}\frac{4A_f\beta 8\pi}{(4\pi)^6} \int_0^\infty dp \, p
  f^f_0(p)[1 \pm f^g_0(p)][\chi^f(p)-\chi^g(k) ]^2\\
  \nonumber
  &\times \left[\frac{2\pi^2}{6}+\frac{\mu^2}{2T^2}-\Dilog \left(\frac{e^{\mu/T}}{1+e^{\mu/T}}\right)-\frac{1}{2}\ln^2\left(1+e^{\mu/T}\right)\right]\\
  &\times \frac{1}{e^{\mu/T}+1}
\end{align}
and
\begin{align}
  \nonumber
  (\chi_{ij},\CC\chi_{ij})^{(D_{\overline{f}})}=& \sum^{\overline{f}h} \frac{4A_f\beta 8\pi}{(4\pi)^6} \int_0^\infty dp \, pf^{\overline{f}}_0(p)[1 \pm f^g_0(p)]\left[\chi^{\overline{f}}(k)-\chi^g(k)\right]^2\\
  &\times \frac{e^{\mu/T}}{e^{\mu/T} + 1}\left(\frac{\pi^2}{6}-\Dilog(-e^{\mu/T})\right)
\end{align}
In each case, the result is suppressed by $e^{-\mu/T}$ in the large $\mu/T$ regime.
Though not obvious from the formulae, this is intuitively clear, since the diagram involves an incoming antiquark, and these are exponentially rare with a fugacity of $e^{-\mu/T}$.

Finally, the Compton scattering diagram (E) differs only slightly from the annihilation diagram (D) in leading log order. As discussed in \cite{Arnold:2000dr} for Compton scattering, the matrix element is given by $(-s/t)$, but at leading order in small $q$ one can consider $-s/t=u/t$. The computation is straightforward and gives:
\begin{align}
  \nonumber
  (\chi_{i...j},\mathcal{C}\chi_{i...j})^{(E_f)}=& \sum^{\fh}\frac{4A_f\beta 8\pi}{(4\pi)^6} \int_0^\infty dp \, pf^f_0(p)[1 \pm f^g_0(p)][\chi^f(p)-\chi^g(p) ]^2\\
  &\times e^{-\mu/T} \frac{e^{\mu/T}}{e^{\mu/T} + 1}\left(\frac{\pi^2}{6}-\Dilog\left(-e^{\mu/T}\right)\right)
\end{align}
\begin{align}
  \nonumber
  (\chi_{ij},\CC\chi_{ij})^{(E_{\overline{f}})}=& \sum^{\overline{f}h}\frac{4A_f\beta 8\pi}{(4\pi)^6} \int_0^\infty dp \,  pf^{\overline{f}}_0(p)[1 \pm f^g_0(p)]\left[\chi^{\overline{f}}(p)-\chi^g(p) \right]^2\\
  \nonumber
  &\times \frac{e^{\mu/T}}{e^{\mu/T}+1}\Big[\frac{2\pi^2}{6}+\frac{\mu^2}{2T^2}-\Dilog\left(\frac{e^{\mu/T}}{1+e^{\mu/T}}\right)\\
  & - \frac{1}{2}\ln^2(1+e^{\mu/T})\Big].
\end{align}
As mentioned at the beginning of this discussion, diagram (E) is also highly suppressed in dense regions.
This occurs because, at leading-log order, the final-state quark's energy equals the initial-state gluon's energy.
The final-state quark must have an energy $E \geq \mu$ to avoid a large Pauli blocking factor, and a sufficiently energetic gluon has a Boltzmann factor of $e^{-\mu/T}$.
So again, the process is suppressed by $e^{-\mu/T}$.
The Compton scattering of an antiquark is similarly suppressed because the incoming antiquark statistical function has the same $e^{-\mu/T}$ suppression factor.
Therefore, processes (D) and (E) not only become less important in comparison with process (C), their reduced importance is exponential and not just polynomial.

\section{Tests and checks}

With the computational tools all set up, we now discuss some internal checks on the procedure.  We also extract some information about which species, and which processes, are the most important as a function of $\mu/T$.
All checks will be conducted on 3-flavor QCD, that is, we take the up, down, and strange quarks to be light and the charm and bottom to be heavy.

\begin{figure}[htb]
    \centering
    \includegraphics[scale=0.70]{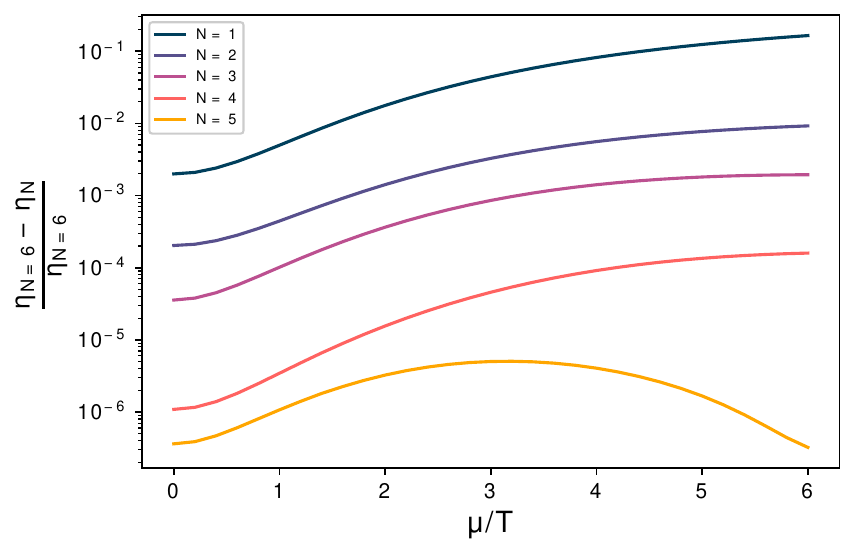}
    \caption{Basis-size dependence of the extracted shear viscosity.
    The curves show the relative error for the different basis set sizes when compared to $N=6$.}
    \label{ratio}
\end{figure}

Our procedure is based on an extremization technique which uses a finite-sized functional basis, as discussed around \Eq{Ansatz}.
We should check how well the results converge as the basis size is expanded.
Therefore we vary $N$ in \Eq{LGYbasis} and examine how the resulting shear viscosity result changes.
This is shown in figure \ref{ratio}, which shows the difference of the $N$-basis answer from the 6-basis answer for $N=1,2,3,4,5$.
We observe a rapid convergence with basis size, although the single-function basis works much less well than in the $\mu=0$ case.
In what follows we will use $N=6$, which the figure indicates is more than sufficient to eliminate issues associated with finite basis size.

\begin{figure}[htb]
    \centering
    \includegraphics[scale=0.70]{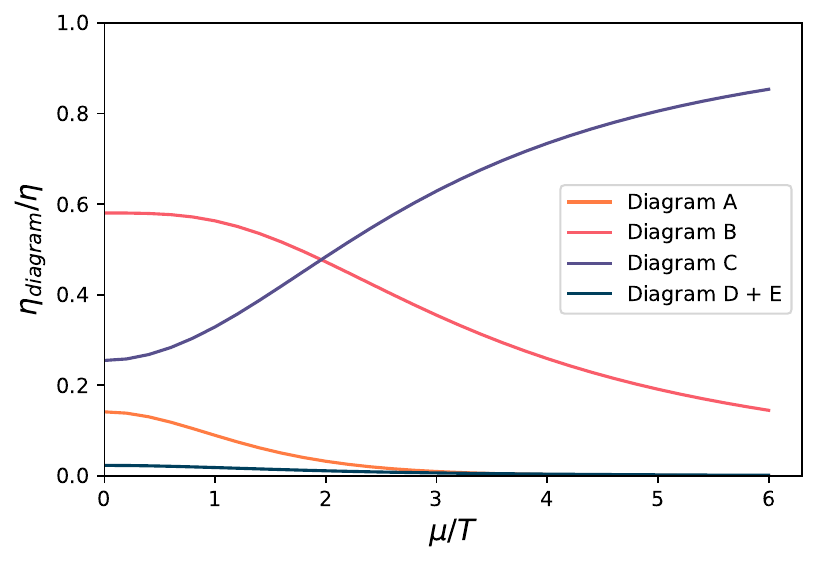}
    \caption{Contribution from each diagram to shear viscosity, $a_m \tilde{\CC}^{\mathrm{(A/B/C\ldots)}}_{mn} a_n$, normalized by the sum of all contributions, as a function of chemical potential.}
    \label{diagrams}
\end{figure}

We claim above that diagram (C) becomes the most important at large values of the chemical potential.
To check this, we first determine the departure from equilibrium coefficients $a_m$ introduced in \Eq{Ansatz}.
Extremizing \Eq{Qbasis}, one finds that
\begin{equation}
    a_m = \tilde{\CC}^{-1}_{mn} \tilde S_n \,,
\end{equation}
which directly gives us the departure from equilibrium for each species.
Inserting this into \Eq{Qmaxval}, we see that the contribution of, say, diagram (B) to the shear viscosity can be expressed as $a_m \tilde{\CC}^{\mathrm{(B)}}_{mn} a_n$.
We show the relative importance of each diagram in figure \ref{diagrams}.
The figure confirms that diagram (C) (quark-quark scattering) dominates at large chemical potential, though (B) (quark-gluon scattering) is the most important until about $\mu = 2T$.
This is roughly the point where quarks start to dominate gluons in setting the Debye screening strength.
The other diagrams are relatively unimportant.
For diagram (A) this is because gluons are closer to equilibrium than quarks, due to their large group Casimir.
For diagrams (D,E) this is already known from Ref.~\cite{Arnold:2000dr}.

\begin{figure}[htb]
    \centering
    \includegraphics[scale=0.70]{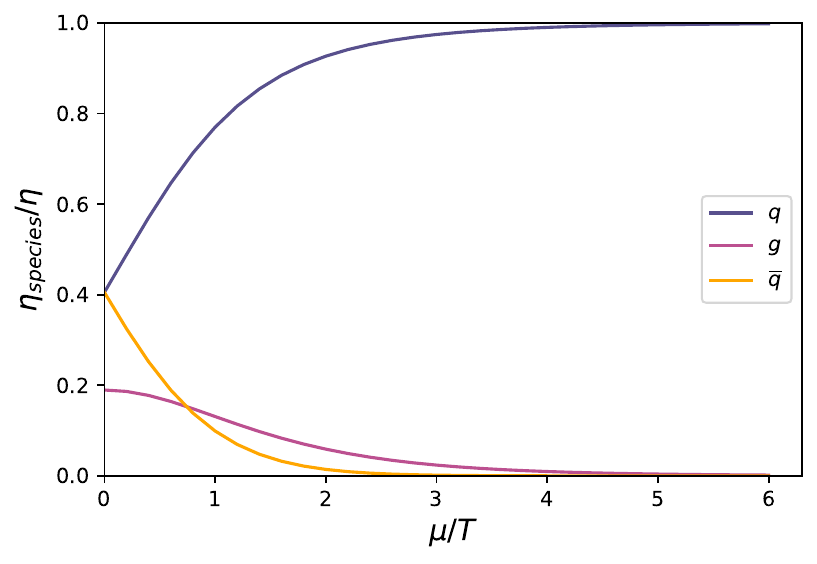}
    \caption{Contribution from each species to shear viscosity, normalized by the sum of all contributions, as a function of chemical potential.}
    \label{Q}
\end{figure}

We can also learn about the relative contributions of quarks, gluons, and antiquarks to the viscosity, by using \Eq{Qmaxval} in the form
$\eta = (\chi_{ij},S_{ij})/15$ and $(S_{ij},\chi_{ij})=a_m \tilde{S}_m$ from \Eq{SCbasis}.
By considering the contributions from quarks, gluons, and antiquarks separately in $a_m \tilde{S}_m$, we obtain figure \ref{Q}.
The figure indicates that gluons are already subdominant at $\mu=0$, and that antiquarks become less important relative to quarks exponentially.
By $\mu/T=2$, quarks completely dominate the shear viscosity, with a small gluon contribution and essentially no antiquarks.

\section{Results}
\label{sec:res}

\begin{figure}
    \centering
    \includegraphics[scale=0.70]{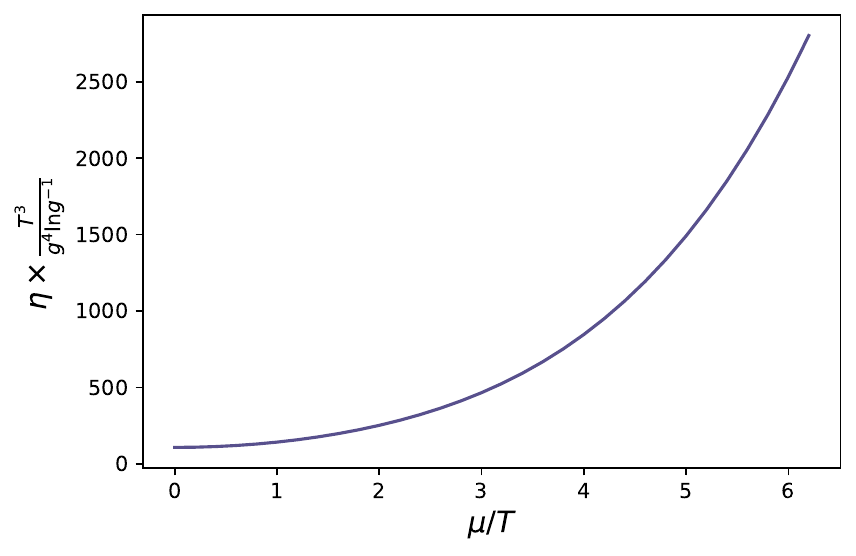}
    \caption{Shear viscosity, normalized to the temperature, for a 6-function basis, 3 flavors of quarks and $N_c=3$, as a function of chemical potential. For $\mu=0$ we recover the value obtained in Ref.~\cite{Arnold:2000dr}.}
    \label{visc6}
\end{figure}

In the last sections, we have shown how one can use the source vector and the scattering matrix to compute shear viscosity as a function of baryonic chemical potential, and also presented a good basis set for these calculations.
Using this computational apparatus, we compute the shear viscosity as a function of $\mu/T$ and display the result in figure \ref{visc6}.

\begin{figure}[htb]
    \centering
    \includegraphics[scale=0.70]{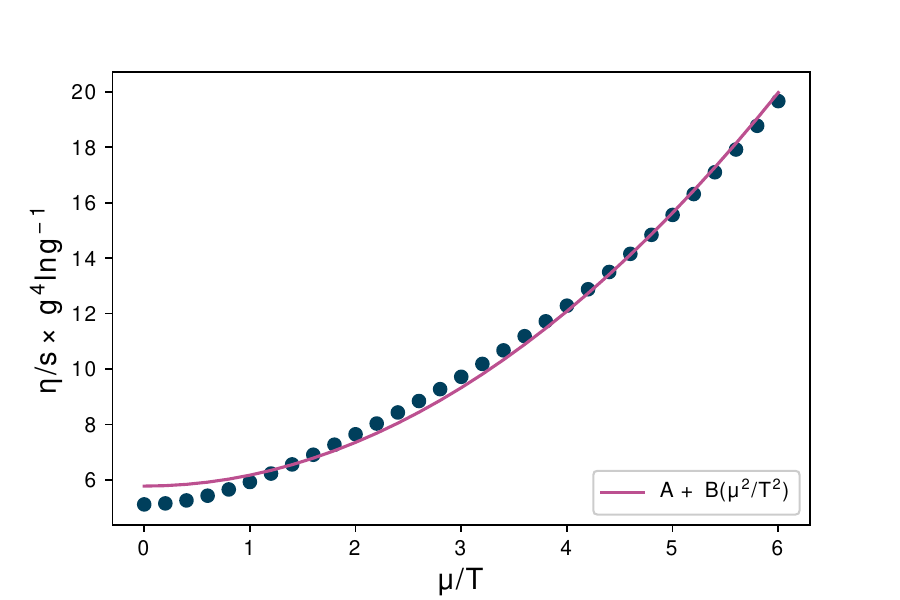}
    \caption{Shear viscosity divided by the entropy density as a function of $\mu$. The points represent our leading-log calculation.
    The pink line is a quadratic fit, which gives a fair but imperfect representation of our result.}
    \label{fig8}
\end{figure}

\begin{figure}[htb]
    \centering
    \includegraphics[scale=0.70]{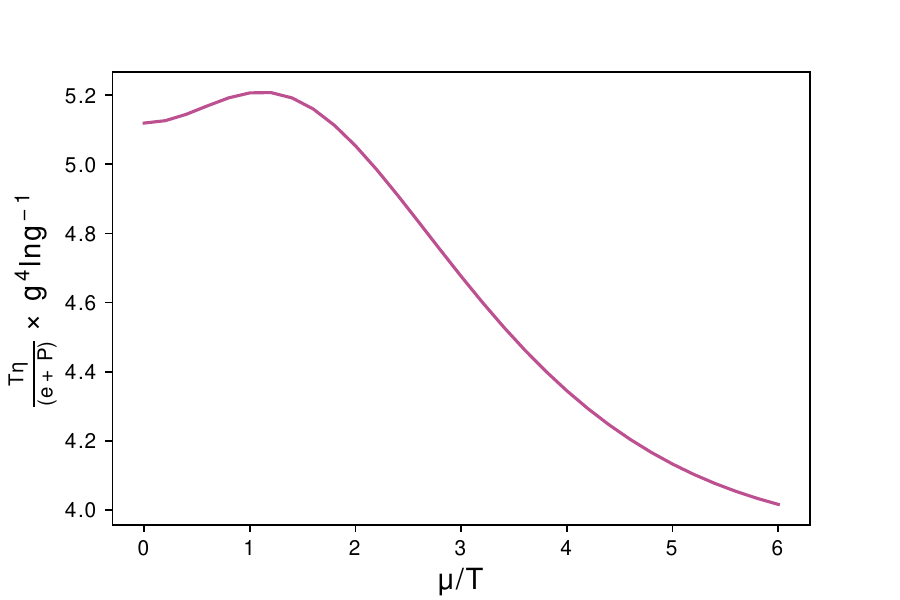}
    \caption{Kinematic shear viscosity $\eta T/(e+P)$ as a function of $\mu$.}
    \label{fig9}
\end{figure}

The figure shows that the shear viscosity depends strongly on $\mu$ when it is expressed in units of the temperature $T$.
Physically, this arises because the number of quarks rises rapidly as $\mu/T$ becomes large at fixed $T$.
Therefore, it makes sense to normalize the shear viscosity in terms of something which gives a more physically relevant result.
In three-flavor QCD, the leading-order entropy density $s$, pressure $P$, and energy density $e$, as functions of $(T,\mu)$, are:
\begin{align}
    s & = \frac{19\pi^2}{9}T^3 + 3\mu^2T
    \label{s}
    \\ \nonumber
    3P = e & = \frac{19\pi^2T^4}{12} + \frac{9}{2}\mu^2T^2 + \frac{9\mu^4}{4\pi^2}
\end{align}
In figure \ref{fig8} we plot the shear-viscosity to entropy density ratio $\eta/s$, and in figure \ref{fig9} we show the kinematic viscosity, that is, the ratio of the shear viscosity to the enthalpy density, $\eta T/(e+P)$.
In each case, a factor of $g^4 \ln(g)$ is extracted such that the result is a single dimensionless curve valid at all coupling strengths in the leading-log approximation, and in each case we use a 6-function basis set and consider the 3-color, 3-flavor case, which is of the most physical interest.

The advantage of $\eta/s$ is that the ratio is directly dimensionless.
This is also a combination which has been speculated to satisfy a conjectured lower bound \cite{Kovtun:2004de}.
Figure \ref{fig8} clearly shows that this ratio follows a parabola in the high-density region.
That can be interpreted as follows.
The dominant scattering mechanism for particles to change their direction is the $\ell(\ell+1) \chi^2$ term in the collision operator in \Eq{ang}. For soft scattering, this is well described in terms of momentum diffusion with a momentum-diffusion coefficient $\hat{q}$:
\begin{equation}
    \hat{q} \equiv \int \frac{d^2 q_\perp}{(2\pi)^2} q_\perp^2 \; C(q_\perp)
\end{equation}
where $C(q_\perp)$ is the differential rate to exchange
transverse momentum $q_\perp$. In a thermal system without chemical potential, this is given by:
\begin{equation}
    C(q_\perp) = g^2 C_F T \frac{m_D^2}{(q_\perp^2)(q_\perp^2 + m_D^2)}
\end{equation}
In a high-density regime the expression for $\hat{q}$ remains the same, but with the Debye screening mass taking the value of $m_D^2 \sim g^2 \mu^2$. This increases the value of $\hat{q} \sim g^4 \mu^2 T $. At the same time, the amount by which a particle must change momentum for the system to equilibrate also increases: $(\Delta p)^2 \sim \mu^2$ rather than $T^2$.  The time scale necessary for the system to equilibrate can then be estimated as:
\[
t \sim \frac{(\Delta p)^2}{\hat{q}}  \sim \frac{1}{g^4 T}
\]
Therefore, shear viscosity can be estimated as:
\[
\eta \sim \frac{P}{t} \sim \frac{\mu^4}{g^4 T} \,.
\]
As \Eq{s} shows, at large $\mu/T$ the entropy scales as $s \propto \mu^2 T$.
Therefore, one expects that:
\begin{equation}
    g^4 \log(g^{-1}) \eta/s = A + B \mu^2/T^2.
\end{equation}
Naturally, the true curve does not follow this form precisely, since there is a transition, as $\mu/T$ is increased, from a system where gluons carry much of the energy and cause much of the scattering to a system where both are dominated by quarks.
However, the expectation for parabolic behavior at large $\mu/T$ is borne out by our calculation.

While $\eta/s$ is of theoretical interest, the kinematic viscosity
$\eta T/(e+P)$ is more physically relevant, because it directly controls the time scale on which the system approaches equilibrium.
This is because the enthalpy density is the quantity which enters the hydrodynamic equations along with the viscosity.
Since both $\eta$ and $e+P$ scale as $\mu^4$ in the large-$\mu$ regime, we expect the ratio $T \eta/(e+P)$ to depend weakly on $\mu/T$ and to approach a constant at large $\mu/T$.
This is indeed what we find.
It is interesting that this constant value is smaller than the value we obtain at $\mu=0$, indicating that, in terms of the time scale $1/T$, a high-density fluid will relax somewhat more quickly than one at vanishing chemical potential.
The same behavior is also present in the QGP phase in calculations of $\eta T/(e+P)$ that were done using holography for strongly coupled fluids\cite{Grefa:2022sav}.

\section{Discussion}

We have performed a first, leading-log, determination of the shear viscosity of QCD at high temperature \textsl{and} baryon chemical potential.
In comparison to the case of vanishing-$\mu$ QCD, we find that the viscosity is larger when expressed in terms of $\eta/s$, but slightly smaller when compared to $(e+P)/T$.
We have also argued that the behavior of Debye vs magnetic screening scales is more favorable at high chemical potentials, and that this implies a wider range of validity for the perturbative analysis.
Though we do not expect the perturbative analysis to work quantitatively down to the temperatures and densities obtained in intermediate-energy heavy ion collisions, we do expect the calculations to work much closer to this regime than proves to be the case in the $\mu=0$ case.
This is because, as we demonstrate, the physics is more dominated by quarks and less by gluons, and the strong mutual interactions between gluons is what leads to large NLO effects on the $\mu=0$ axis \cite{Ghiglieri:2018dib}.
To make this argument concrete, and to provide a higher-quality determination of the shear viscosity, we are encouraged to extend the calculation presented here to leading-order and if possible to NLO.

\section*{Acknowledgements}
The authors acknowledge the support by the State of Hesse within the Research Cluster ELEMENTS (Project ID 500/10.006), and support by the Deutsche Forschungsgemeinschaft (DFG, German Research Foundation) through the CRC-TR 211 'Strong-interaction matter under extreme conditions'– project number 315477589 – TRR 211. The authors also thank Jacquelyn Noronha-Hostler and S\"oren Schlichting for instructive conversations.

\bibliographystyle{JHEP2}
\bibliography{refs}

\end{document}